\documentclass[sigconf]{acmart}
\usepackage{tabularx}
\usepackage{array} 
\usepackage{wrapfig}
\usepackage{graphicx}
\usepackage{subfig}
\usepackage{xcolor}
\usepackage{listings}
\usepackage{color}
\usepackage{float}
\usepackage{xspace} 
\usepackage{multirow}
\usepackage{enumitem}

\AtBeginDocument{%
  \providecommand\BibTeX{{%
    \normalfont B\kern-0.5em{\scshape i\kern-0.25em b}\kern-0.8em\TeX}}}

\copyrightyear{2022}
\acmYear{2022}
\setcopyright{rightsretained}
\acmConference[CHI '22]{CHI Conference on Human Factors in Computing Systems}{April 29-May 5, 2022}{New Orleans, LA, USA}
\acmBooktitle{CHI Conference on Human Factors in Computing Systems (CHI '22), April 29-May 5, 2022, New Orleans, LA, USA}
\acmDOI{10.1145/3491102.3502095}
\acmISBN{978-1-4503-9157-3/22/04}

\begin{document}

\title{Understanding How Programmers Can Use Annotations on Documentation}


\author{Amber Horvath}
\email{ahorvath@cs.cmu.edu}
\affiliation{%
  \institution{Human-Computer Interaction Institute, Carnegie Mellon University}
  \streetaddress{5000 Forbes Ave}
  \city{Pittsburgh}
  \state{Pennsylvania}
  \country{USA}
  \postcode{15213}
}

\author{Michael Xieyang Liu}
\email{xieyangl@cs.cmu.edu}
\affiliation{%
  \institution{Human-Computer Interaction Institute, Carnegie Mellon University}
  \streetaddress{5000 Forbes Ave}
  \city{Pittsburgh}
  \state{Pennsylvania}
  \country{USA}
  \postcode{15213}
}

\author{River Hendriksen}
\email{hendriksenriver@gmail.com}
\affiliation{%
  \institution{University of Pittsburgh}
  \streetaddress{4200 Fifth Ave}
  \city{Pittsburgh}
  \state{Pennsylvania}
  \country{USA}
  \postcode{15260}
}

\author{Connor Shannon}
\email{connorsh@cs.cmu.edu}
\affiliation{%
  \institution{Human-Computer Interaction Institute, Carnegie Mellon University}
  \streetaddress{5000 Forbes Ave}
  \city{Pittsburgh}
  \state{Pennsylvania}
  \country{USA}
  \postcode{15213}
}

\author{Emma Paterson}
\email{emma.paterson@tufts.edu}
\affiliation{%
  \institution{Tufts University}
  \streetaddress{419 Boston Ave}
  \city{Medford}
  \state{Massachusetts}
  \country{USA}
  \postcode{02155}
}

\author{Kazi Jawad}
\email{kjawad@cs.cmu.edu}
\affiliation{%
  \institution{Carnegie Mellon University}
  \streetaddress{5000 Forbes Ave}
  \city{Pittsburgh}
  \state{Pennsylvania}
  \country{USA}
  \postcode{15213}
}

\author{Andrew Macvean}
\email{amacvean@google.com}
\affiliation{%
  \institution{Google}
  \streetaddress{601 N 34th St}
  \city{Seattle}
  \state{Washington}
  \country{USA}
  \postcode{98103}
}

\author{Brad A. Myers}
\email{bam@cs.cmu.edu}
\affiliation{%
  \institution{Human-Computer Interaction Institute, Carnegie Mellon University}
  \streetaddress{5000 Forbes Ave}
  \city{Pittsburgh}
  \state{Pennsylvania}
  \country{USA}
  \postcode{15213}
}

\renewcommand{\shortauthors}{Horvath et al.}

\begin{abstract} 
Modern software development requires developers to find and effectively utilize new APIs and their documentation, but documentation has many well-known issues. Despite this, developers eventually overcome these issues but have no way of sharing what they learned. We investigate sharing this documentation-specific information through \textit{annotations}, which have advantages over developer forums as the information is contextualized, not disruptive, and is short, thus easy to author. Developers can also author annotations to support their own comprehension. In order to support the documentation usage behaviors we found, we built the Adamite annotation tool, which provides features such as multiple anchors, annotation types, and pinning. In our user study, we found that developers are able to create annotations that are useful to themselves and are able to utilize annotations created by other developers when learning a new API, with readers of the annotations completing 67\% more of the task, on average, than the baseline.
\end{abstract}



\maketitle

\section{Introduction}
Application programming interfaces (APIs), including libraries, frameworks, toolkits, and software development kits (SDKs) are used by virtually all code \cite{MyersStylos2016}. Programmers at all levels must continually learn and use new APIs in order to complete any project of significant size or complexity \cite{dualaekoko2012}. In learning APIs, developers depend upon the documentation, including tutorials, reference documentation, and code examples, along with question-and-answer sites like Stack Overflow \cite{Lethbridge2003}. However, there is significant evidence that APIs are often difficult to use \cite{Robillard2009, MyersStylos2016, dualaekoko2012}, which can cause APIs to be used incorrectly, resulting in bugs and sometimes significant security problems \cite{Fahl2013, zhang:2018}. Despite years of research, users still complain about documentation's poor quality, such as the documentation containing ambiguous and incomplete information \cite{Aghajani2019, UddinRobillard2015}, which can severely block users \cite{UddinRobillard2015, roehm2012, RobillardDeLine2011}.

In order to compensate for some of these shortcomings of documentation, we are investigating how \textit{annotations} can be leveraged to help developers more easily learn how to use an API when using its documentation. Annotations are commonly defined as meta-level \textit{notes} that are \textit{anchored} to a specific piece of text (like Microsoft Word \textit{comments}, which are anchored to a point in the document). In a series of studies grounded in developer usage of annotation tools, we investigated ways in which annotations are uniquely poised to address some of documentations' shortcomings while fitting into developers' natural documentation usage and note taking strategies. Using this knowledge, we developed an annotation system that extends the state-of-the-art with documentation-specific features.

The existence of Stack Overflow and other social question and answer sites shows that developers are willing and able to provide content that is helpful to others. Further, Microsoft's Open Publishing model provides evidence that developers will do work to help improve documentation \cite{Microsoft2016}. However, there are known issues with question and answer sites. In a study of what causes Stack Overflow questions to go unanswered, one of the largest factors was a lack of clarity in the question due to lack of context \cite{Asaduzzaman2013}. 

In contrast, annotations provide a tight coupling between the original source of information and the commentary on it, so the context is inherent in where the annotation is anchored. Further, we have identified a collection of new kinds of information which are useful to developers as annotations, but which would \textit{not} be appropriate on a forum such as Stack Overflow due to being highly location-specific and concise. Considering the overhead that goes into writing a question-answer forum post (e.g., \cite{Fourney2013}), annotations are a promising avenue for creating and sharing in-context notes that do not require a large amount effort to author.

As discussed in Robillard and DeLine's field study of API learning obstacles \cite{RobillardDeLine2011}, developers have many questions about the documentation itself that are not easily answered, and annotations can serve as a way of facilitating dialogue among users of the API. Our annotations, especially the ``issue'' type which identifies problems with the documentation or API, can form useful communication between users and key stakeholders, such as documentation writers, who need concise feedback about their documentation \cite{RobillardDeLine2011, PostavaDavignon2004}. There is already evidence that developers take notes on what they have learned \cite{Maalej2009} and find these notes helpful \cite{maalej2014}, so we aim to explore what aspects of these notes, given context by where the annotation is anchored on the documentation, are useful to the developers themselves, as well as to \textit{other} developers.

To explore the concept of annotations as a way of supporting short notes on documentation that are useful both for the author and for later readers, we started with a preliminary lab study that explored the concept of annotations on documentation using an off-the-shelf Chrome extension, Hypothesis \cite{Hypothesis}, and then we performed a corpus analysis of annotations on documentation created using Hypothesis. Given what we learned from these preliminary analyses, we developed our own documentation-specific annotation tool, Adamite\footnote{Adamite stands for \textbf{A}nnotated \textbf{D}ocumentation \textbf{A}llows for \textbf{M}ore \textbf{I}nformation \textbf{T}ransfer across \textbf{E}ngineers and is a green mineral.}. Next, we ran a two-pass user study where we explored the kinds of annotations developers \textit{authored} when learning a new API and then had another set of developers \textit{read} those annotations while attempting to complete the same API learning task using Adamite. We compared these participants to a \textit{control} condition which had no annotations. From these studies, we provide evidence that annotations are useful in helping developers overcome documentation-related issues. 

The contributions of this paper are:
\begin{itemize}
    \item Identifying that developers' note-taking of behaviors, hypotheses, issues, and reminders about the API while using documentation can be facilitated with annotations. 
    \item Identifying that useful information for developers, such as explanations of code, notes about the behavior of the API, and issues in the documentation,  can be provided in the form of short notes as annotations on documentation, where the one or more anchors provide the needed context (Sections \ref{Preliminary Study}, \ref{Corpus Analysis}, and \ref{Lab Study}).
    \item A collection of features integrated into Adamite, including annotation types used to categorize information, pinning to keep track of information, and multiple anchor points for connecting fragmented information, that make annotation authoring and reading more effective when using documentation (Section \ref{Overview of Adamite}).
    \item Identifying prevalent documentation issues that annotations are particularly well-poised to address, including fragmented, ambiguous, incomplete, and incorrect information (Sections \ref{Overview of Adamite} and \ref{Lab Study}).
    \item A study that demonstrates that developers can take notes that benefit themselves through externalizing thoughts and hypotheses about the documentation  (Section \ref{Lab Study}).
    \item A study that demonstrates that developers can take notes that are beneficial to future readers of the documentation (Section \ref{Lab Study}).
    \item Identifying what kinds of notes are beneficial, including code explanations, answered questions, and notes on the behavior of the API, but that hypotheses about the documentation, unanswered questions, and notes that do not build upon the documentation are not beneficial to other developers when learning a new API (Sections \ref{Lab Study} and \ref{Discussion}).
\end{itemize}

\section{Related Work}

Our work builds off three areas of study: programmers learning and usage of API documentation, programmers' note taking behaviors, and annotation systems.

\subsection{Studies of Documentation}

Documentation, specifically API documentation, has been the subject of many research projects, often attempting to understand the particular pain-points of modern software documentation while learning a new API \cite{RobillardDeLine2011, UddinRobillard2015, Lethbridge2003, Aghajani2019, nykaza2002, Meng2019, endrikat2014, roehm2012}, especially considering that API documentation is cited as one of the most important resources but also one of the most significant obstacles when learning a new API \cite{Robillard2009, roehm2012, Chatterjee2017}. In one survey of 323 professional developers \cite{UddinRobillard2015}, incomplete information was the most frequently cited issue with documentation that was a significant blocker to developers. Other highly mentioned blockers included ``ambiguous information'', ``unexplained examples'', and ``incorrect information''. We show that annotations can help with these issues by supporting notes that explain ambiguous information and code examples, while also identifying incorrect information. 
Another study reported that issues with documentation led developers to explore other information sources, such as question-answer sites like Stack Overflow, blog posts, and bug reports, which can contain rich information that may be used to supplement the original API documentation \cite{Chatterjee2017}. We support annotating these other sources and connecting these sources to the original documentation using multi-anchoring, such that this supplemental information can be easier to find.

\subsection{Studies of Programmers Learning in General}

There are many studies of programmers learning unfamiliar code \cite{roehm2012, deline2005, ko2003, maalej2014, xia2018}, some of which are relevant to annotating documentation \cite{dekel2009, rahman2015, Latoza2007, dualaekoko2012, Fourney2013, Sillito2008, Nasehi2012}. LaToza et al. \cite{Latoza2007} discussed that programmers need to learn many facts while understanding code, and would benefit from a way to record what they learned, which was one inspiration for providing annotations as a mechanism for keeping track of information. Another study by Duala-Ekoko et al. found that developers have many questions that they ask when learning unfamiliar APIs that are not trivial to answer by merely reading the API documentation \cite{dualaekoko2012}. When developers have these questions, there is no easy way of attaching an answer to the point that inspired their question -- a problem that our system attempts to address. 

Codepourri \cite{Gordon2015} and code.chats \cite{Oney2018} are annotation-like systems that are designed specifically to try and help with comprehension and discussion around \textit{code}, with both systems showing that in-context discussion of code is effective, further motivating our need to have a similar system for discussing \textit{documentation}. 

\subsection{Studies of Programmers' Note-Taking Behaviors}
Prior work has found that developers write short notes, typically as a way to keep track of and externalize important information \cite{Maalej2009, maalej2014} and to jot down what they are working on when they are interrupted \cite{parnin2010}. One study \cite{maalej2014} found that 40\% of respondents in a survey of 1,477 professional developers used notes as a primary method of comprehending code, and developers that preferred to take notes also typically used their notes when sharing information with other developers. One comprehensive study of developers' note-taking patterns found that developers take different types of notes when comprehending and developing code \cite{cook1993}. Our work extends prior work through a more in-depth analysis of the types of notes that developers make when learning a new API, a process that combines documentation reading with program generation and comprehension, and offers direct tooling support for developers' note-taking needs on documentation.

\subsection{Previous Research on Annotations and Annotation Systems}
A prior literature review \cite{Agosti2007} found that the flexible nature of annotations allows them to serve a variety of purposes, including supporting in-context commenting and creating connections among parts of text. Other work noted that annotations may be seen as a conversational tool among the document users, as well as with the document creators \cite{Fogli2004, Guzdial2000} -- a property that is useful for documentation writers who need feedback \cite{RobillardDeLine2011, PostavaDavignon2004}.

Other systems that support annotating on the web helped inspire and inform our design. Chilana et al.'s system LemonAid allows users to select web page elements such as buttons and menu items and crowd-source questions about the element's intended usage and answers about each element \cite{Chilana2012}. Zyto et al. developed an annotation system called NB which was successful in math and science classrooms \cite{Zyto2012}. Other systems \cite{Fang2021, vermette2017, hong2009} have used sharing of annotated materials to assist in learning online materials, but none of these systems were focused around specifically trying to improve the underlying annotated document or assist programmers.  A commonly-used annotation system is Hypothesis, a browser extension that supports annotating text on web pages \cite{Hypothesis}. We utilized Hypothesis for our preliminary evaluations of using annotations in documentation.

\section{Preliminary Study}\label{Preliminary Study}
To explore the efficacy of annotations as a useful learning device for API learning tasks, we ran a preliminary study where people learned an unfamiliar API while using Hypothesis \cite{Hypothesis}. In summary, we found that developers are able to author annotations in the ways we envisioned, but that annotation authoring and reading could be improved for developers by adding additional tooling features.

\subsection{Hypothesis}
Hypothesis \cite{Hypothesis} is a browser extension that supports annotation creation and reading using a sidebar that is fixed to the side of the browser window. Users can create an annotation by highlighting text on the page, which will cause a pop-up to appear -- users may choose to either add an annotation with text or simply highlight the selected text. When choosing to create an annotation with text, the sidebar will update with a rich text editor. Users may publish their annotations publicly, privately, or to a group. Once an annotation has been published, the text on the web page will be highlighted in a light yellow color. Users may reply to annotations that have been created, share an annotation using a hyperlink, or flag an annotation for moderator review if the content is deemed offensive. 

\subsection{Design}
The preliminary study had two distinct phases: the first phase was focused on understanding how developers \emph{author} annotations during an API learning task, while the second phase focused on how developers \textit{read} annotations that are already attached to documentation, even when there are many irrelevant annotations.

Participants were instructed to complete some Python code using Apache Beam (an API learning task adapted from a previous study \cite{Horvath2019}), while foraging through Beam's documentation for the requisite information. This task was chosen because it is difficult and requires understanding the documentation which has previously-reported short-comings \cite{Horvath2019b}. All participants were recruited from the authors' social circles, had 45 minutes for the task, and had some experience with Python -- 4 participants were recruited for the first phase, and 5 participants were recruited for the second phase. Task instructions for each condition and the starting code can be found in the Supplementary Materials.

In the \textit{authoring} condition, participants were given the Beam documentation with no annotations, and instructed to add annotations when they learned anything useful, had questions about the content in the documentation, or had any other thoughts about the documentation. Participants could also annotate other websites, such as Stack Overflow, with annotations related to the task. 

In the \textit{reading} condition, participants were given the same Beam documentation, but with annotations added. We provided all the annotations, which were of two types --- annotations authored by the first author that were designed to be helpful for this task given what developers were confused about in a previous study \cite{Horvath2019}, and other annotations that were designed to be ``distracting'' to simulate the more realistic case where not all annotations would be relevant. We consider these annotations ``distracting'' in that they are not related to the task the participant is trying to complete. We collected our ``distractor'' annotations from a number of Stack Overflow posts that were related to Apache Beam and chose a subset of 44 question-answer pairs that were in Python and were relatively concise and understandable. These question-answer pair annotations were anchored to Beam's documentation on words or phrases that matched the text in the original question. In total, we had 23 ``helpful'' annotations and 44 ``distractor'' annotations, totalling 67 unique annotations. 

\subsection{Preliminary Study Results and Discussion}\label{Preliminary Study Results}
In the authoring condition, the 4 participants together authored a total of 19 unique annotations. Each participant, on average, authored 4.75 annotations, with annotations averaging 4.41 words. Hypothesis also allows users to simply highlight a piece of text on a web page without adding any text content to the anchor (hereafter referred to as ``highlight'' annotations) --- out of the 19 unique annotations authored, 5 were these simple highlight annotations.

Many of the annotations that participants created showed a part of the documentation that illustrated how to achieve some part of their current task. Other annotations served as task reminders or open questions the author had about the documentation content. 

Annotations adapted from Stack Overflow were not particularly helpful, as evidenced by their distracting nature deterring participants from reading other annotations. This was caused in part because the distractor annotations were too long, thus participants struggled to determine their relevance. The length also made it difficult to determine why the annotation was anchored in its particular location and what it had to do with the documentation's content. These results suggest that annotations in documentation must be shorter in length and have information that is highly relevant and focused around the anchor content. 

In our analyses, we also found that participants had many questions about the documentation (on average, 10.8 questions per participant) which were not annotated. While trying to answer these questions, participants routinely encountered more confusing information, resulting in them losing track of their original questions. Given this confusion, participants struggled to answer their questions, with only 29\% of questions definitely answered. Notably, Hypothesis does not have any way of marking or following up on a question.

From this initial preliminary study, we found evidence that annotations may enhance the original text. Additionally, we found support for different types of information needs that are not directly supported by Hypothesis, such as keeping track of open questions. We also learned that annotations need to be easy to skim and relatively short in length. Since participants had trouble finding annotations that met their needs, we also found a need for better filter and search support, along with a need to anchor annotations in multiple places so that developers can more effectively find annotations when reading documentation. 

\section{Corpus Analysis of Hypothesis Annotations}\label{Corpus Analysis}
In order to supplement our preliminary study, we queried Hypothesis's API to get a list of public annotations which developers have already made on official API documentation including public APIs from Google, Microsoft, Oracle, and Mozilla, along with other developer learning resources including Stack Overflow, W3Schools, and GitHub. 

Across these sites, we found 1,995 public annotations\footnote{This count does not include private annotations, so this count is most likely only a subset of all of the annotations made on these sites.}. Of the 1,995 annotations, 196 were questions about the content of the documentation, and 995 were highlight type annotations. An additional 16 annotations expressed confusion without specifically being a question (e.g., ``I don't understand this''). Of the 1,000 annotations with content, 16 of the annotations were to-do items the author wanted to follow up on, 43 pointed out problematic aspects of and potential improvements to the documentation, and 79 were created to specifically call out important or useful parts of the documentation\footnote{These counts were generated by counting instances of phrases like ``incorrect'' and ``todo'' in the annotation content, then manually reviewing all of the annotations that contained those phrases to determine if they were actually referencing an issue, helpful part of the document, or todo item.}. These annotations were authored by 298 unique users (average = 6.694 annotations per user, minimum = 1, maximum = 677) across 1,143 unique web pages. The authored annotations were, on average, 8.79 words long and were anchored to text that averaged 12.35 words. 

We believe many of the 1,995 annotations that contained content could benefit other developers with additional tooling support. For example, a Hypothesis user annotated the text ``you can pass the path to the serve account key in code'' and asked how they can do that. This user then later annotated a code example showing how to achieve this behavior at a different point in the documentation and said ``finally found it''. While these two annotations depend upon one another in order to make sense and point to different parts of the documentation, Hypothesis does not allow for these annotations to reference one another, suggesting a need for better tooling support for multiple anchors for annotations and keeping track of open questions. 

These annotations provide support for our claim that some developers are willing to write annotations and attach them to documentation, as they are already doing this. Moreover, the annotations we found follow some of the patterns we identified in our preliminary study, such as open questions and issues. However, Hypothesis's general-purpose annotation system does not have enough support to effectively utilize these annotations.

\begin{figure*}[htbp]
\centering
 {\includegraphics[width=\textwidth]{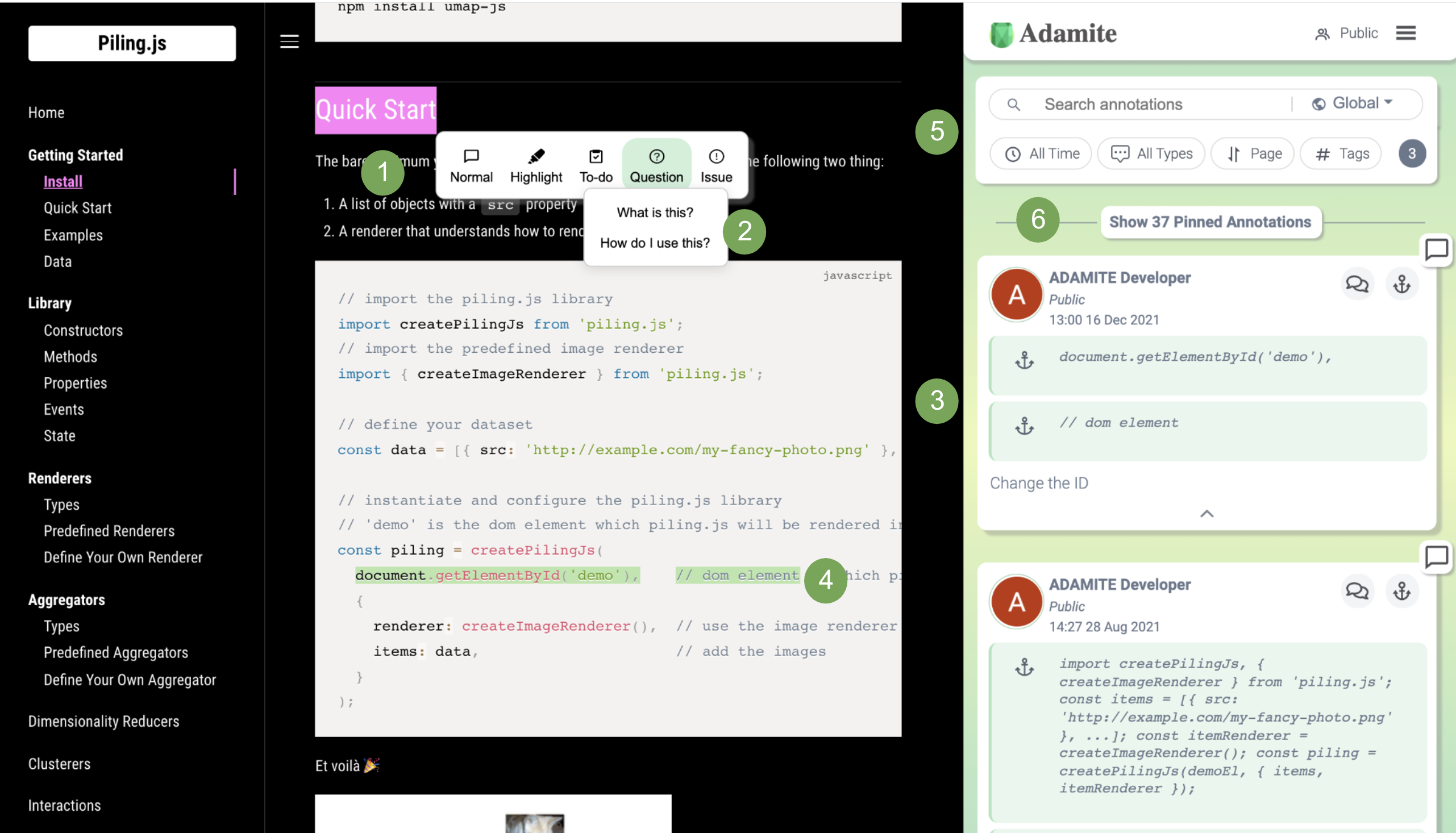}}

\caption{Adamite's sidebar (on the right) open on an already-annotated web page in the browser. (1) shows the pop-up for when a user selects some text -- at this point they can begin creating a new annotation by selecting an annotation type. (2) shows the menu of question annotation prompts users can choose from. (3) shows a published normal annotation with two anchors. (4) shows how the annotated text appears on the web page. (5) shows Adamite's search and filter pane. (6) shows the pinned annotation list button.}
\label{Adamite Piling}
\end{figure*}

\section{Overview of Adamite}\label{Overview of Adamite}
We designed Adamite, a browser extension, specifically to help developers keep track of important information, organize their learning, and share their insights with one another. We intentionally designed Adamite to not only support features that can address known documentation issues, but also to support the developer who is making the annotations (see Table \ref{featuretable}). To create an annotation, a developer, who we call the annotation ``author'' simply needs to open the Adamite sidebar, highlight some text on the web page (called the ``anchor''), select the type of annotation, add text to a rich text editor that appears in the sidebar, and click on the publish button. Once published, the text that the user annotated will be highlighted on the web page and the annotation will appear in the Adamite sidebar -- see Figure \ref{Adamite Piling}. Users may also add tags and additional anchors to the annotation. Annotations can be published publicly, privately, or to a group of Adamite users. Once an annotation has been published, it may be replied to by others, and edited or deleted by the original author. Clicking on the anchor icon on the annotation will scroll to the part of the web page the annotation is anchored to (or will open a new tab if the anchor is on a different page) -- conversely, clicking on the highlighted text on the web page will scroll to the corresponding annotation in the sidebar.

One goal of Adamite is helping developers structure and share what they learn in the documentation in a way that is useful both to themselves and for later developers. To achieve this, we developed annotation types. In addition to the typical ``normal'' (with a user-written comment) and ``highlight'' (just the anchor and no comment) annotations, Adamite supports question, issue, and to-do annotations. We chose these three annotation types to assist developers in keeping track of their questions, to point out and possibly attempt to rectify issues found in the documentation, and to help them keep track of their tasks. Issue annotations have a button intended to alert key stakeholders, such as the documentation writers, of the described problem with the documentation. Question annotations are stateful, meaning unanswered annotations will stay available until the developer either marks the question as ``answered'' (at which point the answer will be appended to the original question), or marks the question as ``no longer relevant''. To-do annotations are also always available until they are marked as complete. 

Question and to-do annotations are always available using Adamite's ``pinning'' mechanism. Most annotation systems only show annotations that are on the user's current web page. However, considering that documentation may be spread across many pages and developers may visit many web pages when attempting to complete a programming task, we added in the ability to \textit{pin} an annotation, such that is always available in a list at the top of the sidebar. To-do and question annotations are pinned by default, since the developer is unlikely to find their answer or finish their task while they are on the same web page.

Given that a common documentation problem is fragmented information, we found a need to support \textit{multiple} anchors for a single annotation. This feature can be used to connect parts of the documentation that the user feels should be presented together, or to better contextualize their annotation. Developers may also use anchors as a way of collecting multiple parts of the documentation that they feel are related to one another given the developer's task and their evolving understanding of the API. 

For later users of the annotated documentation (who we call annotation ``readers,'' but can be the same person as the annotation authors), it is likely that not all of the annotations are relevant to what the developer is trying to do. To help readers find the most relevant annotations, we support search (using Elasticsearch \cite{elasticsearch}) and filters (see Figure \ref{Adamite Piling}-5). Readers can search across a web page, website, or across all of Adamite's annotations and readers can filter on the annotation type, when the annotation was created, and what tags the author has tagged the annotation with. Readers may also sort the annotations by their location on the page or by the time at which the annotation was authored.

\begin{table*}[t]
\centering
\resizebox{1\textwidth}{!}{%
\begin{tabular}{
 >{\raggedright}p{.2\textwidth} | >{\raggedright}p{.7\textwidth} 
}

 \toprule
  \textbf{Feature} & 
  \textbf{What problem/behavior is this addressing?}  \tabularnewline 
  \midrule
 
  Annotating & 
  Developers sometimes take notes on what they have learned \cite{Maalej2009, maalej2014, parnin2010, liu2019} 
  \tabularnewline
  \midrule

  Question Annotation & 
  Developers have many questions about unfamiliar APIs and their documentation (\cite{dualaekoko2012, Sillito2008}, Sections \ref{Preliminary Study Results} and \ref{Corpus Analysis}) 
  \tabularnewline
  \midrule

  Issue Annotation & 
  Identify documentation issues including obsoleteness \cite{UddinRobillard2015, Aghajani2019}, incorrectness (\cite{UddinRobillard2015, Aghajani2019}, Section \ref{Corpus Analysis}), incompleteness \cite{UddinRobillard2015, Robillard2009, RobillardDeLine2011, ChenHuang2009, Aghajani2019, Zhi2015, Cummaudo2020} and ambiguities \cite{UddinRobillard2015, Robillard2009,  RobillardDeLine2011, Aghajani2019} 
  \tabularnewline
  \midrule

  To-do Annotation & 
  Developers take notes on open tasks that they must work on, especially when interrupted \cite{parnin2010}, and occasionally take notes on tasks they must complete that are related to parts of the documentation they are reading (Section \ref{Corpus Analysis})
  \tabularnewline
  \midrule

  Multi-Anchoring & 
  Developers need to build up a mental representation of an API \cite{kittur2013, Horvath2019b, liu2019} and connect related resources \cite{Cummaudo2020, Aghajani2019, UddinRobillard2015} 
  \tabularnewline
  \midrule

  Search and Filter & 
  Developers, especially selective \cite{gendermag} and opportunistic \cite{Brandt2009} learners, want to quickly find information that is relevant to them (\cite{Meng2019}, Section \ref{Preliminary Study Results}) 
  \tabularnewline
  
\bottomrule

\end{tabular}
}
\caption{\label{featuretable} Adamite's feature set and how it relates to previously-reported developer needs.
}
\end{table*}

\subsection{Implementation}\label{Implementation}
Adamite is a Chrome extension built using React 16.13, and Webpack 4, with annotations and user profile data stored in Firestore \cite{firestore}. Upon every page load, the extension makes a request to the Firestore database for that web page's annotations that the user has access to. The annotations are continually updated on that web page using Firestore's \texttt{onSnapshot} event listener which listens for any change to the annotations that were received through the query, in case other users add annotations to the open web page. For search, we utilize ElasticSearch \cite{elasticsearch} as Firestore does not natively support text-based searching.

For anchoring, Adamite goes through the annotations that are on the current page and uses the annotation's stored XPath to find the part of the web page to highlight. We generate the XPath using a recursive XPath building function that converts each nested DOM element into an XPath. We additionally store a copy of the text that the user annotated, along with the starting offset and ending offset which are the character counts from the beginning of the element and the end of the element to the text, respectively. If the page's content changes and the text that the XPath finds does not match the text that is saved, our algorithm traverses up the DOM to try to find a matching string close by, and if not, then we mark the anchor as broken and allow the user to fix it (but see Future Work).

\section{Lab Study}\label{Lab Study}
In order to understand the role that annotations play in developers' documentation usage while learning a new API, we ran a lab study with three conditions to understand how developers create and use annotations. Participants in one condition \emph{authored} annotations while completing an API learning task, and participants in the second condition \emph{read} these participant-authored annotations. The third condition was a \textit{control} condition where participants completed the same API learning task using just the documentation. The lab study consisted of a training task, a programming task, and a survey to assess the participant's background -- all study materials, including the files necessary for the programming task, are available in the Supplemental Materials. The study was approved by our institution's Institutional Review Board.

\subsection{Method}
\subsubsection{Training}
Each condition included a training exercise using Tippy, a React library for making tooltips, and its documentation to either familiarize the participants with Adamite and its functionality (Adamite conditions) or to familiarize them with thinking aloud while reading through documentation (control). Participants in the Adamite conditions learned how to create an annotation of each type, reply to an annotation, add an additional anchor to an existing annotation, search, filter, edit and delete an annotation and practiced thinking aloud while performing these tasks. The control condition practiced thinking aloud when they had a question, found an answer to their question, and identified an issue in the documentation.

\subsubsection{Task}
For the task, participants were asked to complete an image aggregation and organization task using Piling.js (hereafter referred to as ``Piling''), a JavaScript library for handling visualizations \cite{piling}. Piling was chosen as it is a relatively small library, meaning the participants would have adequate time to gain a high-level understanding of the library and its functionalities during a lab study. Further, Piling is a relatively unknown library, thus the documentation is particularly important as it is one of the few sources of information on the library.

\begin{figure}
\begin{centering}
\includegraphics[width=.475\textwidth]{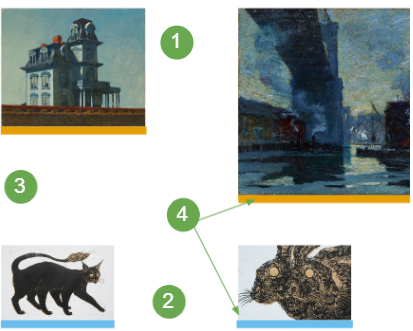}
\end{centering}
\caption{The correct output for the task. Each number refers to the step number. (1) creates and renders the 4 images. (2) puts the images in 2 rows. (3) arranges images by a user-defined property using the arrangeBy method. (4) requires the user to set a label on their data, such that elements with the same label will have a matching stripe along the bottom of the picture.}
\label{TaskImage}
\end{figure}

The task was to use Piling to take a set of four provided images and render and sort the images (see Figure \ref{TaskImage} for the output and detailed steps). The task was chosen as, despite its apparent simplicity, it requires the participant to learn some of Piling's core concepts including how Piling's rendering works, how to set properties in Piling, and how to structure and refer to data that is passed into Piling. Participants were objectively graded upon how many of the 4 steps they were able to complete correctly. To start, participants were given a JavaScript file containing comments stating the goal of each step.

In addition to completing the programming task, participants were asked think aloud and to pretend as though they were in a small team learning Piling. Dependent upon the condition, further instructions differed slightly. \textit{Control} condition participants were told that they needed to relay what they had learned to their teammates in whatever way they would normally do so, whether that be through notes or some other mechanism. Adamite \textit{authoring} participants were instructed to create annotations with any questions or thoughts they had about the documentation, issues they found in the documentation, and thoughts they wanted to follow up on and that these annotations would be shared with their future teammates. Each authoring participant started with an un-annotated version of the documentation. In the Adamite \textit{reading} condition, participants were given annotated documentation and were told to pretend that the annotations were created by a teammate who had already learned Piling and were instructed to speak aloud when an annotation was helpful or unhelpful. Participants in the reading condition were not required (but were allowed) to create annotations or interact with the annotations present in the documentation. 

\begin{figure}[htbp]
\begin{centering}
\includegraphics[width=.475\textwidth]{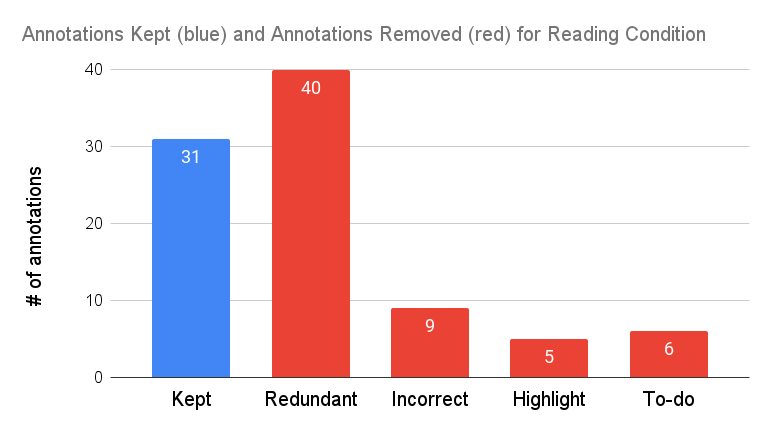}
\end{centering}
\caption{The number of annotations removed for each reason, along with the annotations kept, out of the 91 total annotations. 2 highlight annotations were retained as the users edited them to add text, making them semantically identical to normal annotations.}
\label{Annotations Kept}
\end{figure}

\subsubsection{Annotation Selection}
In choosing annotations to include in the reading condition, two researchers separately coded each annotation created during the authoring condition for whether or not to include it. Inclusion criteria included identifying matching annotations across participants to select which one was the clearest, most appropriately anchored, and concise -- qualities informed by our preliminary study and others \cite{Agosti2007}. Notably, the chief cause for the majority of annotations to be removed was redundancy -- participants commonly annotated information related to the first two steps of the task (see Figure \ref{Annotations Kept})\footnote{Note that this is due to being a lab study -- in a realistic situation, people would likely read an existing annotation and not create a redundant one.}. We also excluded annotations that the participant later stated were incorrect or that the participant deleted, and annotations that lacked sufficient context (including all highlight annotations). Finally, we omitted to-do annotations, as they are designed only for the original author's usage. The researchers had a 71\% agreement --- in the cases where the researchers disagreed, they had a discussion until agreement was reached. Through this process, we were left with 31 annotations. One additional annotation was added by the first author to assist with the final step of the task, as no participant in either the authoring condition or control condition got to that point of the task. After this process, we had 32 annotations, with 18 normal type annotations, 10 issue type annotations, and 4 question annotations, 3 of which were answered\footnote{We included one unanswered question to account for the realistic situation that not all questions would be answered and because the question asked was a common question among participants -- notably, answering this question was not necessary for succeeding in the task.}. We did \textit{not} omit any annotations due to relevance or correctness, since we wanted to leave in anything that at least one participant wanted to comment on to be more realistic.

\subsubsection{Participants}
We recruited 31 participants using departmental mailing lists at our university, the authors' social circles, and advertisements about the study on Twitter and a Reddit forum for JavaScript developers. One participant could not finish the study due to technical difficulties, so we only report on the 30 who completed the whole study. Each condition included 10 participants and were randomly assigned between the authoring and control condition -- the reading condition occurred after the other two conditions so all remaining participants who signed up were assigned to that condition.

All of the participants were required to have some amount of experience using JavaScript, not to have used Piling before, and to have been programming for at least 1 year (actual minimum: 1 year, maximum: 20 years, average: 7.98 years). The participants' professions included graduate students in computer science-related fields, user experience researchers with a computer science background, and professional programmers. The gender breakdown of our study consisted of 19 men, 9 women, and 1 non-binary person. Participants across each condition had a similar amount of JavaScript experience and years of programming experience. 

All study sessions were completed remotely using video conferencing software. Participants were audio and video recorded, and each participant's session took approximately 90 minutes, with 45 minutes of that time allotted for the programming task. Each participant was given access to the Piling documentation and a CodeSandbox.io \cite{codesandbox} project which had JavaScipt, HTML, and CSS files with Piling installed and a photo of the output, along with written-out steps for the task. Participants were compensated \$25 for their time, save for 2 participants who elected not to be compensated.

\subsubsection{Analysis Methods}
Across all of the conditions, we objectively graded participants on whether or not they succeeded in completing each of the 4 steps outlined in the task instructions. In the Adamite conditions, we analyzed the video recordings and log data to count how many annotations participants authored, and how often they filtered, searched, clicked on anchors, pinned, replied to, edited, read, revisited, or deleted their annotations in order to understand how developers integrated annotating into their workflow. Reading and revisiting an annotation was coded objectively by only counting an annotation as read or revisited if they expanded the annotation\footnote{Annotations were collapsed by default, meaning only a preview of the content was visible and expanding the annotation required clicking on the annotation -- we took this as an indication that the participant was interested in its content.} or read aloud its content. Creating an annotation was not counted as reading or revisiting so some annotations have counts of 0.

We qualitatively coded the annotations developers made in order to characterize developers' annotating strategies. Using an open coding method, two authors coded the normal type annotations by independently coding each annotation and refining categories based upon their individual codes. For issue and question type annotations, we coded the annotations dependent upon what issue in a list of commonly defined issues was identified in the annotation (issue type) or what issue caused the participant's confusion (question type, see Table \ref{Issue and Question Annotations}). Two of the authors independently coded the annotations and reached 75\% agreement when coding the issue annotations and 73\% for the question annotations -- remaining annotations were discussed until agreement was achieved. 

In the annotation reading condition, we analyzed how often participants said that an annotation was helpful or unhelpful in order to better understand what annotations succeeded in helping participants. We objectively coded this through only marking an annotation as helpful or unhelpful if a participant explicitly stated this during the think-aloud. We calculated average helpfulness by how many participants said an annotation was helpful and dividing by how many participants encountered the annotation. We ranked annotations from most helpful to least helpful by how many participants said the annotation was helpful subtracted by how many said it was unhelpful\footnote{We chose to rank annotations this way such that we could account for the impact of an annotation -- if an annotation helped 7 people and did not help 1 person (i.e., 7/8 = 87\% helpfulness), we did not want that to be seen as ``less helpful'' than an annotation that helped the only participant to encounter it (i.e., 1/1 = 100\% helpfulness).}. 

In the control condition, we kept track of whether and how the participant chose to relay their information to their teammates. We also referenced the auto-generated transcripts to find and count whenever a participant stated a question. 

\begin{figure}[htbp]
\begin{centering}
\includegraphics[width=.475\textwidth]{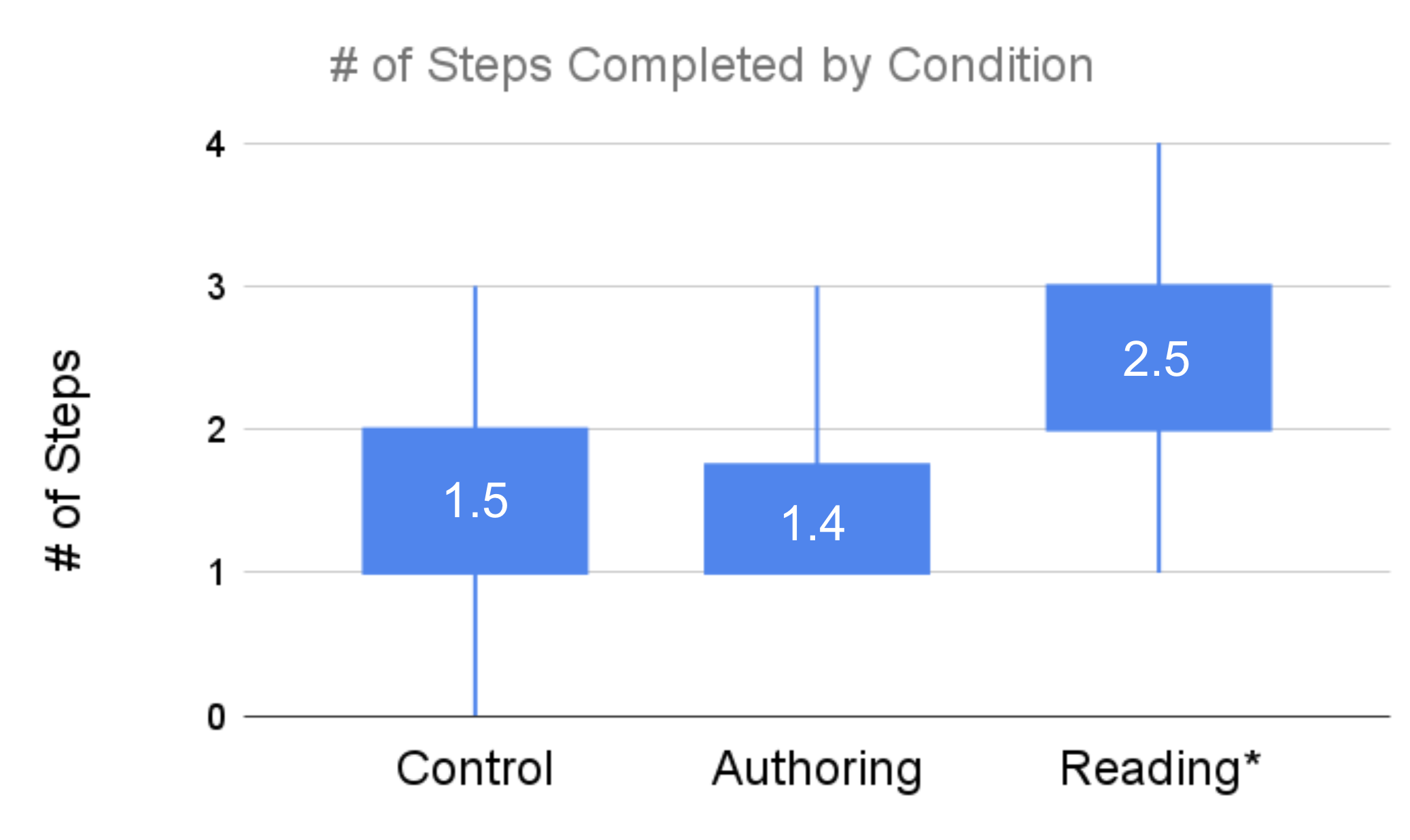}
\end{centering}
\caption{The difference between reading and each of the other two conditions is statistically significant, but the difference between control and authoring is not. The boxes represent the range of steps completed between the first and third quartiles per condition, and the lines represent the minimum and maximum number of steps completed per condition. The average number of steps completed is in the center of each box.}
\label{Steps Completed}
\end{figure}

\begin{table}[t]
\centering
\begin{tabular}{
>{\raggedright}p{.14\textwidth}|
>{\raggedright}p{.07\textwidth}|
>{\raggedright}p{.071\textwidth}|
>{\raggedright}p{.09\textwidth}
}
\toprule
  \textbf{Documentation Issue} & \textbf{Issue Annos.} & \textbf{Question Annos. About Issue} & \textbf{Percent Questions Answered} \tabularnewline \midrule [0.1ex] 
 Incompleteness & 3 & 6 & 50\% \tabularnewline \midrule
 Fragmentation & 4 & 3 & 33\% \tabularnewline \midrule
 Incorrectness & 4 & 2 & 0\% \tabularnewline \midrule
 Poor Code Example & 10 & 7 & 43\% \tabularnewline \midrule
 Ambiguity & 2 & 12 & 33\% \tabularnewline 
 \bottomrule
\end{tabular}

\caption{\label{Issue and Question Annotations} Counts of each issue and question annotation that identified or was caused by an issue discussed in \cite{UddinRobillard2015}. Note that two code examples did not work because they suffered from a fragmentation issue, so they are coded both as a poor code example and a fragmentation issue. Similarly, all fragmentation questions were caused by fragmented code examples, so they are also coded as both a fragmentation question and code example question.}
\end{table}

\subsection{Results}
On average, participants in the control completed 1.5 of the 4 steps, authoring participants completed 1.4 steps, and the reading condition completed 2.5 steps (see Figure \ref{Steps Completed}). Participants in the reading condition performed significantly better than participants in the control and authoring conditions (paired T-test versus control, \emph{p < .01}, paired T-test versus authoring, \emph{p < .01}). In the control condition, 1 participant chose to take notes in a Google Doc and 1 participant made comments in their code as notes for their future teammates. 2 participants spoke aloud when they had a thought that they would want to share as a note to their teammates, but did not actually write any notes down. The remaining 6 control participants did not take notes or verbally indicate the intent to take notes at any point during the study. This suggests that without a mechanism for externalizing their thoughts, these 6 participants may not have been able to actually share what they learned and only 2 participants had any artifact to share with their future teammates.

\begin{figure}[htbp]
\begin{centering}
\includegraphics[width=.475\textwidth]{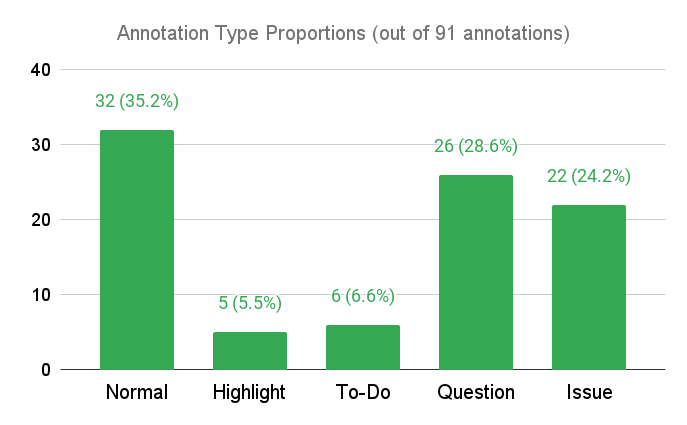}
\end{centering}
\caption{The proportions for each type of the annotations made in the authoring condition (out of 91), with the exact count for each type above the bar and the proportion in parentheses. }
\label{Annotation Counts}
\end{figure}

\begin{figure*}[htbp]
\begin{centering}
\includegraphics[width=.9\textwidth]{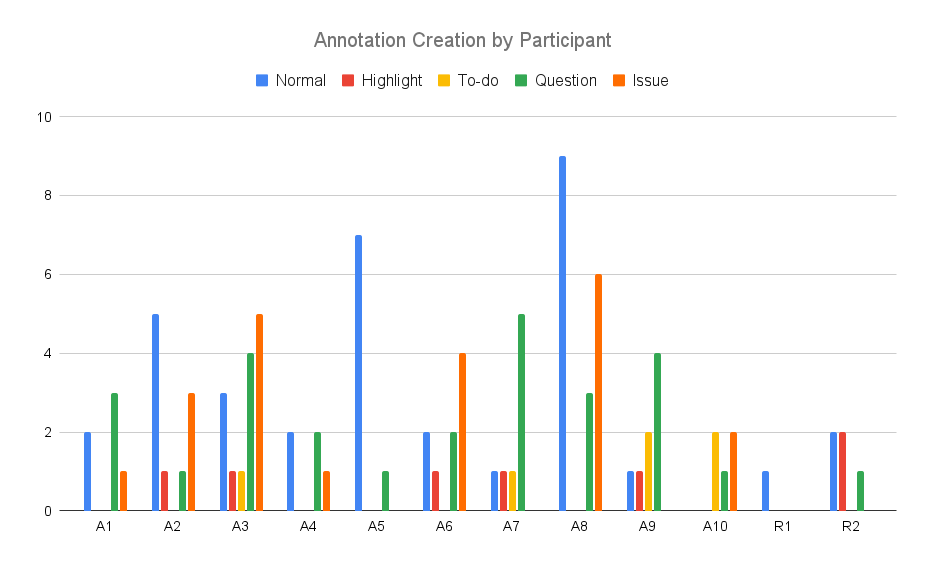}
\end{centering}
\caption{The breakdown of which participants made what types of annotations. A1 through A10 refer to the 10 authoring condition participants. R1 and R2 are the two reading condition participants who created annotations.}
\label{Annotation Creation}
\end{figure*}

The 10 participants in the annotation authoring condition created 91 annotations across all five of the annotation types (see Figures \ref{Annotation Counts} and \ref{Annotation Creation}). On average, each participant authored 9.1 annotations (median = 8, standard deviation = 4.094, minimum = 5, maximum = 18), with the most used annotation type being the normal-type at 32 authored annotations (35.1\%). 2 participants in the annotation \textit{reading} condition created 6 annotations (3 normal, 2 highlights, and 1 question), resulting in 97 annotations across all conditions. The rest of the analyses just look at the 91 annotations from the authoring condition.

\begin{table} 
\centering
\begin{tabular}{
>{\raggedright}p{.14\textwidth}|
>{\raggedright}p{.07\textwidth}|
>{\raggedright}p{.071\textwidth}|
>{\raggedright}p{.09\textwidth}
}
 \toprule
  \textbf{Annotation Category} & \textbf{Num. Annos.} & \textbf{Avg. Times Revisited} & \textbf{Num. Retained for Reading Condition} \tabularnewline [0.1ex] 
 \midrule
 Note to Self & 12 & 1.8 & 6  \tabularnewline \midrule
 Explanation of Code & 10 & 0.3 & 7 \tabularnewline \midrule
 Hypothesis & 7 & 0.66 & 3  \tabularnewline \midrule
 Important to Task & 2 & 0.0 & 2 \tabularnewline\midrule
 Other & 1 & 4 & 0 \tabularnewline
 \bottomrule
\end{tabular}
\caption{\label{Normal Usage} Counts of each coded normal annotation, how many of each type were retained for the reading condition, and how often the 10 annotation authors revisited their annotations. Some annotations were revisited more than once.}
\end{table}

\subsubsection{Notes Developers Take When Learning a New API}
Considering the large amount of normal type annotations and how normal annotations can contain nearly any type of information, we sought to characterize the content of these annotations. Through open coding, two coders defined 5 categories -- ``note to self'' in which the participant made a note about the documentation's content that was primarily for themselves, ``explanation of code'' in which the participant tried to better explain what a particular code example was doing, ``hypothesis'' in which the participant guesses about how some part of Piling works, ``important to task'' in which the participant highlights a particular part of the documentation as critical for one of the steps of the task, and ``other'' for any annotations that did not fit into the previous categories. With this categorization, we had 12 ``note to self'' annotations, 10 ``explanations of code'',  7 ``hypotheses'', 2 ``important to task'' annotations, and 1 ``other'' annotation (see Table \ref{Normal Usage}). The 1 ``other'' annotation was an annotation with no content that was created purely as a navigational aid. 

The ``note to self'' annotations typically served as reminders to the author to externalize an important detail about the API or a code example. For example, one participant annotated a call to \texttt{document.getElementById('demo')} with ``remember to change the ID'' as a reminder to themselves, as they were in the process of adapting the example. This note could also benefit future users of the documentation as a note that the code example will not work without some modifications. 

Unexplained or poorly explained code examples are a frequent problem in documentation \cite{Aghajani2019, UddinRobillard2015} and Piling was no exception, so our participants attempted to explain some of the code examples and, sometimes, contextualize them to the goals of the task. The most helpful and second most helpful annotations are both explanations of code with the most helpful explaining how to use Piling's row property to create columns, and the second most helpful annotation explaining how the code example for \texttt{piling.arrangeBy} works and how to adapt the code example to work using a callback function -- both necessary steps for completing the task.

Participants also hypothesized about parts of Piling, including how the library worked and whether the various constructs were relevant to the task. One participant annotated a code example that used an undefined parameter and said ``I think that k is equal to the number of photos in the data set. I could be wrong though - TBD''. Another participant was trying to determine what function to use to sort their images and annotated \texttt{piling.groupBy} with the text ``This might be helpful'', but, upon finding \texttt{piling.arrangeBy}, annotated that method with ``Actually, maybe this''. These hypotheses along with ``note to self'' annotations demonstrate how annotating can be a lightweight technique for jotting down thoughts as a developer is gaining familiarity with an unfamiliar library.

Annotations that marked parts of the API that were important to the task often called out a particular part of the documentation that had necessary information. For example, one participant annotated the text ``Properties'', a heading in the documentation, and said ``This is a table of properties'' -- whereas this seems redundant with the content, it served as a prominent navigation aide.

Considering roughly half of the authored normal annotations are primarily beneficial to the original author (i.e., notes to self and hypotheses) and every participant made a personal annotation (i.e., notes to self, hypotheses, to-do's, and highlights), we find evidence that annotating is an effective mechanism for externalizing information and helpful for the author. We also included 6 notes to self and 3 hypotheses in the reading condition to see whether these thoughts could benefit other developers. Notes to self, in particular, were the most revisited type of annotation by their authors, and participants, on average, revisited these notes 1.8 more times -- more than any other annotation type or coded normal annotation types.

Some of these normal type annotations were used in conjunction with Adamite's other novel features, resulting in the annotations being more useful. 4 of the normal type annotations contained multiple anchors, and 4 others were pinned by participants, resulting in 8 of the 32 (25\%) normal annotations utilizing one of Adamite's novel features for annotating. The most commonly revisited annotation, a note to self, had 5 anchors with each anchor describing a necessary step in order to properly instantiate the \texttt{piling} object. The participant pinned this annotation such that they could reference the anchor steps in their CodeSandbox project (which was open in a separate Chrome tab) -- this annotation was also useful in the annotation reading condition with one participant replying to thank the author. This shows that Adamite's features not only support the creation of lightweight notes but also allow developers to utilize their and other people's notes in context.

\subsubsection{Questions and Issues Developers Annotate}
All authoring condition participants created at least one question annotation and 7 out of 10 authoring participants created at least 1 issue annotation. 3 participants used question type annotations more than any other type, and 2 participants used issue type annotations more than any other type, suggesting the inclusion of these types was helpful (see Figure \ref{Annotation Creation}). 

Issue-type and question-type annotations accounted for roughly half of all the authored annotations. Nearly all issue annotations succeeded in identifying at least one of the issues identified in \cite{UddinRobillard2015} (see Table \ref{Issue and Question Annotations}), save for one issue annotation that stated that a particular part of Piling is ``super high maintenance for a simple use case'', which is not an issue with the documentation, but with the library itself. Notably, this issue annotation was the third most helpful annotation in the reading condition with participants appreciating that it warned them about a part of the library they were thinking of using, suggesting that issue annotations are useful beyond identifying documentation problems. 

Poor code examples were the most frequently identified issues (Table \ref{Issue and Question Annotations}). These code examples typically did not work either because a variable in the example was undefined and thus the code could not be just copy-pasted (7/10) or because the documentation did not show an output of what the code example actually did (3/10). Some of the undefined variable issues occurred because the variable was defined in a different part of the documentation (2/7) -- a documentation fragmentation issue as well as a code example issue. One participant was able to use multiple anchors to suggest where the definition for the variable should be moved to in order to make the code example work.

Participants succeeded in answering 10 of their 26 question annotations, resulting in 38\% of questions being answered (average = 49\% of questions answered across participants, median = 36\%). Considering the task is difficult and the documentation has many issues, the relatively small amount of questions answered is unsurprising. However, 4 participants were able to answer all of their question annotations, providing evidence that participants can answer their open questions when they are annotated with tooling support, such as that the annotation remains pinned as a reminder to revisit it. Sure enough, participants revisited their question annotations fairly often, on average revisiting 1.23 times, suggesting that, when a developer had a question, they made an attempt to follow-up on it. 

Developers also had many questions that relate to documentation issues reported by prior studies \cite{UddinRobillard2015, Aghajani2019}. Ambiguity and poor code examples were the source of the majority of developers' questions, which matches the findings reported in \cite{UddinRobillard2015}, with ambiguity, in particular, standing out as a common and severe blocker for developers. Ambiguity and fragmentation issues also resulted in questions that were difficult for annotation authors to answer, with only 33\% of questions caused by ambiguity and 33\% of questions caused by fragmentation being answered. Considering some participants were able to solve fragmentation issues using multiple anchors with Adamite, this suggests these developers' questions may have been answered if they had been presented with these annotations. In fact, in the Adamite reading condition, 2 participants had their issue of \texttt{aggregateColorMap} not compiling solved by an annotation that used multiple anchors to link to the part of the documentation that defines \texttt{aggregateColorMap}.

\subsubsection{Developers' Annotating Behaviors}
The annotations that participants authored were, on average, relatively short in length at 9.31 words (minimum = 0, maximum = 34, median = 8). The short length of annotations makes them relatively easy to author. The annotations included in the reading study averaged 11.94 words and the 10 most helpful annotations were, on average, 13 words long. These results suggest that short notes are able to help future users of documentation, while not requiring a large amount of effort on the author's part to create. These annotation lengths are also consistent with the annotations authored using Hypothesis in our corpus analysis, suggesting these annotations are similar to the types of annotations authored in the wild.

Considering annotating may be costly in terms of time and effort, we were interested in understanding the benefit the annotations have for the authors themselves. We assessed usefulness for the author through counting how often participants revisited an annotation, pinned an annotation, and how often they clicked the anchors on the annotation as a way to navigate the documentation (see Table \ref{Author Usage}). As shown in the table, the authors were able to utilize their annotations in the ways we intended. All participants revisited at least one of their annotations at least once (min = 1, max = 36, average = 10.225 revisits per participant), with question and normal type annotations revisited most often. These results suggest authors were able to utilize their annotations for their own benefit.

In terms of task completion, participants in the annotation authoring condition on average completed 1.4 steps out of the 4 steps required to complete the task (standard deviation = 0.69, minimum = 1, maximum = 3, median = 1) (see Figure \ref{Steps Completed}). There is no statistically significant difference between the control condition completion rate of 1.5 and the 1.4 in the authoring condition (two-tailed T-test, \emph{p = 0.78}), suggesting that annotating the documentation, while not increasing their performance, also did not require so much overhead that participants were unable to complete the task in the same amount of time as if they had not been annotating. Moreover, considering how often developers' revisited their notes, specifically their ``notes to self'', this suggests authors were able to successfully use annotations as an externalization of their thoughts.

\begin{table*}[t]
\centering
\begin{tabular}{
>{\raggedright}p{.15\textwidth} 
>{\raggedright}p{.1\textwidth} 
>{\raggedright}p{.1\textwidth} 
>{\raggedright}p{.08\textwidth} 
>{\raggedright}p{.1\textwidth} 
>{\raggedright}p{.1\textwidth} |
>{\raggedright}p{.1\textwidth} 
>{\raggedright\arraybackslash} p{.15\textwidth} 
}
 \toprule
  & \textbf{Normal} & \textbf{Highlight} & \textbf{To-Do} & \textbf{Question} & \textbf{Issue} & \textbf{Total} \tabularnewline [0.1ex] 
 \midrule
 Times Revisited & 36 & 5 & 4 & 32 & 12 & 89 \tabularnewline
 Times Edited & 5 & 3 & 0 & 4 & 1 & 13 \tabularnewline
 Times Pinned & 5 & 0 & n/a & n/a & 0 & 5  \tabularnewline
 Anchor Clicks  & 10 & 0 & 0 & 8 & 5 & 23 \tabularnewline
 
 \bottomrule
\end{tabular}
\caption{\label{Author Usage} Counts of how often the 10 annotation authors interacted with their annotations by type. Some annotations were revisited more than once. Question and to-do annotations are pinned by default, so we do not count how many times they are pinned.}
\end{table*}

\subsubsection{How Developers Use Annotated Documentation}
Participants in the reading condition read, on average, 23.7 annotations (including revisiting annotations they had already read, with 72\% of annotations read more than once), and read, on average, 15.6 unique annotations. Participants found 45\% of the annotations that they encountered helpful, and only 8\% not helpful. The top-performing 6 participants in the reading condition also reported the highest proportion of helpful annotations, suggesting that their success may be attributed to the successful use of the annotations.

Of the 32 annotations included in the reading condition, the most helpful type of annotation for readers of the documentation was answered question annotations, with, on average, 54\% of participants who encountered them stating they were helpful. Normal-type annotations were the second-most helpful type of annotation (avg. 47\% helpful) and issue type annotations helped on average 35\% of the time. Some issue type annotations were more helpful than other issue annotations including annotations that identified poor code examples, which were helpful, on average, 45\% of the time. Even though these issue annotations did not necessarily suggest a solution, they did work in confirming the participant's suspicions that the documentation itself was incorrect and not the participant's implementation. Sometimes, participants found useful annotations through search -- participants searched a total of 78 times and 11 of these searches returned an annotation that the user immediately found useful.

Participants especially appreciated the normal type annotations that were explanations of code, with participants finding them helpful 63\% of the time. Code explanations typically elucidated what a code example was illustrating or explained how to adapt a particular code example for the purposes of the task. ``Notes to self'' were also surprisingly useful, with participants finding them helpful 53\% of the time -- given that the notes to self typically represented a thought or reminder the developer had about the documentation while completing the task, these results suggest that the participants in the reading study had similar thoughts about the documentation. Conversely, hypotheses were not very helpful, with only 16\% of participants finding them helpful -- given the uncertainty of these annotations, participants may have found them less trustworthy. These results suggest that explanations of code and developers' personal notes can be useful if they are framed in a knowledgeable fashion, while hypotheses are more useful for the original author.

Two participants in the reading condition chose to annotate, creating a total of 6 new annotations. One participant, who completed 3 steps in the task, was working on the last step of the task and made 2 ``note to self'' and 2 highlight annotations to keep track of and navigate to important parts of the documentation. The other participant made a new annotation, a ``hypothesis'', about what argument was needed for Piling's \texttt{arrangeBy} method. These annotations suggest that, even when using already-annotated documentation, personal annotations may still be useful.

Three reading participants chose to reply to and pin annotations in the documentation. Participants replied to thank authors, build upon the annotation with more specific information, ask for clarification, and confirm that the annotated issue in the documentation is a problem, suggesting users of documentation are able to improve annotations. Pinned annotations were used to keep track of useful annotations other users had left with one participant using their pinned annotation as a quick link to an important part of the documentation, suggesting pinning is not only useful for the original author, but also for later users.

\section{Discussion}\label{Discussion}
Our results suggest that annotated documentation is useful for documentation readers in overcoming some of the known barriers of documentation and that the act of annotating when learning a new API can help developers keep track of their thoughts and open questions. Creating annotations was also useful to the author as a form of self-explanation, which has been shown to be useful for learning in prior studies \cite{chi1989self, chandler1991cognitive}, and these self-explanations, or ``notes to self'', were useful to others. The novel features of Adamite, especially types, multiple anchors, and pinning, helped annotation authors better structure and contextualize their information and helped annotation readers find relevant information.

Participants particularly enjoyed that the annotations had types, and also envisioned future enhancements. 7 participants in the authoring condition noted that they enjoyed the question-type annotations, with 2 specifically mentioning the two built-in question menu items, suggesting that assisting in annotation authoring may be a fruitful avenue for future annotation systems. One participant made an issue type annotation, but wanted the issue to only be shared with documentation writers, while 2 other annotators wanted the ``issue'' type annotation to be less ``confrontational'' and instead frame the annotation as a ``suggestion'' to the documentation maintainers. 
 
Having types for annotations also resulted in two completely separate classes of annotations users made. As demonstrated in our qualitative coding, the kinds of information developers noted in their normal annotations (i.e., notes to self, important to task, hypotheses, and explanations of code) is very different from the information that developers noted in their issue and question annotations, which were primarily documentation-focused. This suggests that annotation typing is an effective way to elicit information through annotations that may not otherwise be noted. 

The most helpful annotation was anchored to the text ``columns'' and simply states ``Use this to create rows'' --- a short, 5 word annotation that explains how this property can achieve an effect required by the second step of the task that is not immediately clear when reading the documentation. The second most helpful annotation also succeeded in elucidating how to use part of the API that is relevant to the task through using multiple anchors and clarifying a code example for \texttt{arrangeBy} -- a method necessary for the third step. In the reading condition, participants were more successful in completing these two challenging steps, with 9 participants able to complete step 2 and 4 participants able to complete step 3. This increase in performance suggests that participants were able to utilize what the annotation authoring condition learned in order to more effectively complete their task.

Annotations that were not as immediately relevant to the task could also be helpful. Two issue type annotations were the third and fourth most helpful annotations, each warning participants about unhelpful and incorrect parts of the documentation. For example, the fourth most helpful annotation, which helped 4 participants, stated that a code example in the documentation throws an error that a variable is not declared --- participants found this annotation useful as it deterred them from using that code example or, if they did use it, reassured them that they were not doing something wrong, since another user had the same problem.

Conversely, the \textit{least} useful annotations were the ones that lacked enough context to be reusable. For example, an unhelpful annotation was an annotation anchored to the text ``columns 10'', stating that the default value of \texttt{column} is 10, which is redundant with the text of the anchor. The original author annotated this as the reason their 4 images showed up in a single row since the column parameter needs to change to make 2 rows, however, the annotation is missing this full context. Future annotation systems designed to help programmers should explore automatically inferring additional context to make the annotation more comprehensible to later users --- if Adamite were to be integrated with the developer's integrated development environment (IDE), we may be able to capture the code and its output before and after the user created the annotation to better explain why they made the annotation and what they were trying to achieve. Communicating additional context about the first user's behaviors and goals to later users who are completing a programming task has been shown to be an effective approach \cite{liu2021}.

In the creation and evaluation of Adamite, we sought to explore to what extent annotations may help annotation authors and readers in overcoming previously-reported shortcomings of documentation. Through this exploration, we have evidence developers are able to identify documentation issues using annotations and are able to answer some of their documentation questions. Specifically, our participants were able to identify ``incompleteness'' and ``ambiguity'' issues, two of the largest blockers when using documentation \cite{UddinRobillard2015}. Other developers can make use of these answered questions and issue annotations, with answered questions as the most helpful annotation type and explained code examples also helping annotation readers. However, annotations cannot solve every documentation issue. If the API and its documentation are updated, the annotations may go out of date, at which point they may be more harmful than helpful. While our algorithm attempts to reattach the annotation to its anchor point, the annotation content will not change to reflect that reattachment, at which point the content may be incorrect. Future versions of Adamite should investigate how documentation writers and the original annotation authors should manage their annotations if they go out of date with an API update and notify annotation authors when this occurs. 

Notably, Adamite and annotations in general are not appropriate in every developer learning situation. Annotations are generally short in length -- prior literature found private annotations averaged 2 to 10 words and public annotations averaged 30 to 150 words \cite{Marshall2004} -- for longer form information, such as a tutorial explaining how to use the API, annotations may not be appropriate. Annotations also are contextualized to their web page and anchor point -- they are meant to be discovered in-context, so, if they are anchored to text that is rarely encountered, they may be less useful as they are unlikely to be discovered. To assist in discovery, we can imagine extending Adamite's search capabilities to be smarter by querying on metadata such as automatically-added tags dependent upon the annotation or documentation's content.

Adamite's feature set and annotations, as a whole, may be appropriate in other domains. When learning a new API, a developer must forage for information while forming a mental model of the API and how it relates to their task, with much of the information they encounter potentially being irrelevant. Other tasks, such as planning a trip or figuring out what type of camera to buy, follow similar patterns where the user must try and ascertain what is relevant to them and learn what is or is not important when attempting to make a decision \cite{kittur2013}. Annotating may be a useful mechanism for keeping track of important information with the annotation serving as rationale for why this information was thought to be important and the anchor can serve as a link back to the original web page and its content that the user found to be relevant. Some of Adamite's features may be effective for these tasks, such as using multiple anchors to link together multiple pieces of related information that serve as rationale for the user's ultimate decision.  However, some features of Adamite may need to be modified -- for example, participants liked the feature to add system-provided questions, but some of Adamite's current automated questions such as ``how do I use this'' may make less sense in non-programming contexts.

\section{Limitations}
Given Piling's complex documentation, Adamite may not be as helpful when the documentation is simpler or clearer, so the study cannot necessarily be said to apply to those situations. Our lab study was also constrained to a single forty-five minute session, so it is unclear how developers' API learning and annotation authoring and reading behavior would change over a longer period of time. Piling also has a small user-base, so we have less evidence that Adamite would be useful for APIs with better documentation or APIs with a large user-base that can provide useful crowd-sourced information on Stack Overflow or mailing lists. Future work should see how developers use Adamite in the wild with more popular APIs over a longer period of time. 

Considering we selected the annotations to be included in the reading condition, there is an additional limitation that this curating process would not happen in the wild, and, since the last annotation added was created by a researcher, we cannot say that every annotation was participant-authored. The most common reason for removing an annotation was due to the annotation's content being redundant with another annotation which would be less likely to occur in the real world where users can see other users annotations and will most likely be performing different tasks. While annotations unrelated to the user's task may be distracting, some annotations, such as issue annotations, may be useful to any developer using the construct(s) the annotation references. Further, Adamite supports tagging and filtering which could be used to filter out annotations that are unrelated to what the user is working on. Annotations could also be curated in order to ensure higher quality annotations are more commonly seen using crowd-sourcing methods (e.g., supporting voting and editing other users' contents \cite{Allahbakhsh2013}) which could be added to Adamite.

Adamite as a tool is also limited by its inability to work on dynamic web pages such as Google Docs since dynamic web pages do not have stable anchor points for our highlighting algorithm and the content on these web pages often changes, causing the annotation to lose its original context. Considering developer documentation is relatively static, Adamite works well in this situation, but we do not claim that Adamite will work on more volatile pages. Adamite also does not work on PDFs, despite some documentation existing in the form of a PDF. API documentation is a good use case for Adamite, though, since there are many well-known issues with documentation that annotations can address and API documentation is commonly presented on a website. 

\section{Future Work}\label{Future Work}
We are interested in further understanding annotations’ role in the software learning process. Currently our annotations are localized to the web browser, but a large part of software learning occurs in the IDE as programmers are exploring and understanding unfamiliar libraries. We plan to extend Adamite’s annotating capabilities to an IDE such that developers can provide code examples straight from the IDE and annotate their own code to more directly and contextually support the note taking needs found in our study and prior work \cite{maalej2014, Maalej2009, parnin2010} while improving code understanding \cite{chandler1991cognitive}. One motivation for annotating code is that prior work has found that developers have many hypotheses about what the code does \cite{Latoza2007} --- in these situations, question type annotations may be a useful mechanism for keeping track of those parts of the learning task and their eventual answers may be useful to other developers. Cook et al. \cite{cook1993} also found that developers take notes on program behavior, such as tracing variable value changes during debugging, which an annotation tool may assist with.

One limitation of Adamite being a browser plugin is that developers cannot revisit or follow up on their annotations if they are in situations where they do not have access to a desktop web browser, such as on a phone. Further, some operations that act on annotations as a batch which we hypothesize may be useful, such as mass deleting or mass tagging, would be more effective through a web site. Batch operations can assist in the often cumbersome task of ``clean-up'' when a user has created a lot of notes, with only some of the notes still being useful. We have developed a prototype version of this website, which may be seen in the supplementary video, but work remains in order to make it more useful. We also plan to improve Adamite's anchoring algorithm using fuzzy anchoring \cite{fuzzyanchoring} such that more annotations remain viable for longer as the annotation attempts to re-anchor itself when the page changes, thus lowering the amount of annotations that need to be ``cleaned-up''.

Lastly, our work is limited by the fact that it was completed in a lab study, which can not fully encapsulate real-world usage of an annotation tool for developers, such as how a developer's usage of Adamite may change as they gain more familiarity with an API and attempt to complete different tasks with it. We are planning on addressing this limitation through running a field study where small teams may use Adamite for a longer period of time.

\section{Conclusion}
Poor API documentation is a known barrier when attempting to learn a new API. Our preliminary study, development of Adamite, and the user study together provide evidence that annotations can be beneficial for helping to mitigate some of the well-known shortcomings of API documentation, while also providing additional benefits such as helping developers keep track of their thoughts and questions through short, in-context notes. When using Adamite to author annotations, developers were able to answer their open questions, point out problematic aspects of the documentation with suggestions for improvement, and create annotations that are useful to themselves and were later also useful to other developers. In particular, Adamite's annotation types and multiple anchors helped developers better contextualize their information, even when the annotation content was short. When reading annotations, developers were able to use these annotations while learning difficult concepts in order to more efficiently complete a programming task.

\begin{acks}
This research was funded in part by the NSF under grant CCF-2007482 and by gifts from Google. Any opinions, findings, and conclusions or recommendations expressed in this material are those of the authors and do not necessarily reflect those of the sponsors. In addition, we would like to thank Aniket Kittur, Imtiaz Rahman, Lai Wei, our participants, and our reviewers for their thoughtful feedback.
\end{acks}

\bibliographystyle{ACM-Reference-Format}
\bibliography{chi22-275-adamite}

\end{document}